\documentclass[aip,apl,groupedaddress,a4paper]{revtex4}

\usepackage[english]{babel}
\usepackage{amsfonts}
\usepackage{amsmath}
\usepackage{amssymb}
\usepackage{amsthm}
\usepackage{dsfont}
\usepackage[dvips]{graphicx}
\usepackage[dvips]{hyperref}

\renewcommand{\phi}{\varphi}

\newcommand{\bra}[1]{\langle #1|}
\newcommand{\ket}[1]{|#1\rangle}
\newcommand{\braket}[2]{\langle #1|#2\rangle}

\begin{document}
\title{Security analysis of the time-coding quantum key distribution protocols}
\author{Thierry Debuisschert, Simon Fossier}
\affiliation{Thales Research and Technology, Campus Polytechnique, 1 avenue Augustin Fresnel, 91767 Palaiseau Cedex}

\begin{abstract}
We report the security analysis of time-coding quantum key distribution protocols. The protocols make use of coherent single-photon pulses. The key is encoded in the photon time-detection. The use of coherent superposition of states allows to detect eavesdropping of the key.
We give a mathematical model of a first protocol from which we derive a second, simpler, protocol. We derive the security analysis of both protocols and find that the secure rates can be similar to those obtained with the BB84 protocol.
We then calculate the secure distance for those protocols over standard fibre links. When using low-noise superconducting single photon detectors, secure distances over 200 km can be foreseen. Finally, we analyse the consequences of photon-number splitting attacks when faint pulses are used instead of single photon pulses. A decoy states technique can be used to prevent such attacks.
\end{abstract}

\maketitle

\section{Introduction}
\label{sec:Introduction}

Quantum key distribution (QKD) exploits the fundamental principles of quantum mechanics to securely distribute a cryptographic key between two parties usually called Alice and Bob. The purpose of QKD is not to prevent a third party Eve from eavesdropping the line, but to make eavesdropping systematically detectable by Alice and Bob.
According to their information advantage over Eve, Alice and Bob can distill a secret key. Quantum key distribution has been widely developed in recent years~\cite{gisin:rmp,scarani-2008}. The proposed protocols are based on photon-counting~\cite{BB84,Ekert1991,B92,PlugAndPlay1997} or continuous variable as well~\cite{grosshans:prl02,grosshans:nature,Lodewyck2007,Weedbrook2004}. An important effort has been devoted to the realization of practical and reliable prototypes~\cite{Stucki2002,Stucki2005,Fossier2009,Treiber2009,Yuan2007}
that has culminated recently with the demonstration of fully integrated field demonstrations of QKD networks within the framework of the SECOQC project~\cite{secoqc,peev2009secoqc} and of the UQCC 2010 conference~\cite{UQCC}.

\begin{figure}
\vspace*{13pt}
\includegraphics[width=0.6\columnwidth]{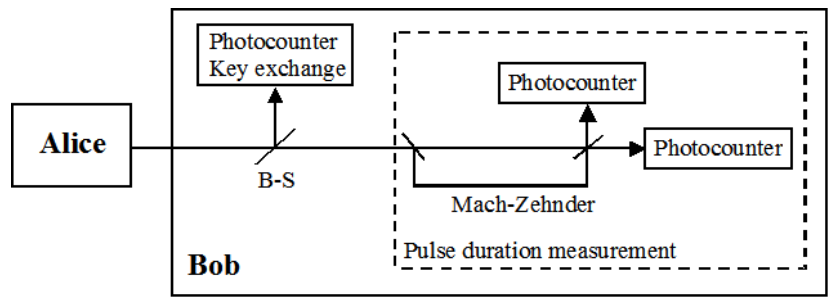}
\vspace*{13pt}
\caption{Typical scheme of the experiment. Bob measures at random the time-detection of the photon or the coherence of the pulse. Here, a beamsplitter sends the pulses either to a photon-counter or to a Mach-Zehnder interferometer.}
\label{TypicalScheme}
\end{figure}

\begin{figure}
\vspace*{13pt}
\includegraphics[width=0.6\columnwidth]{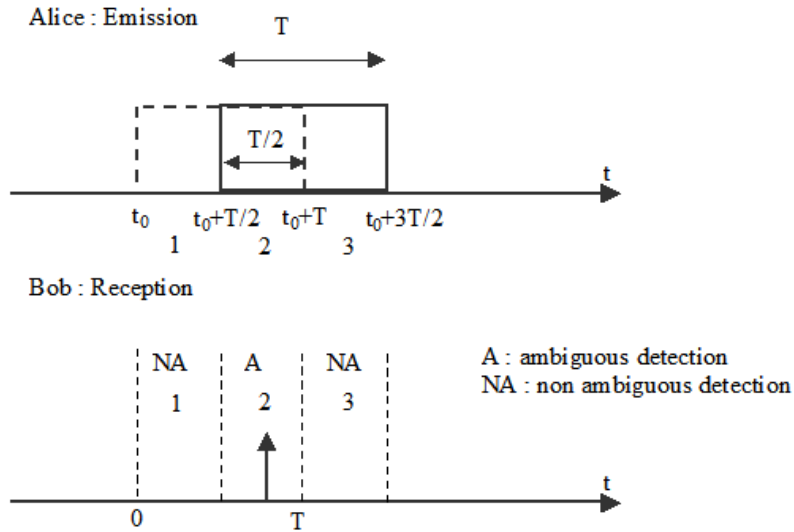}
\vspace*{13pt}
\caption{Principle of the Three Time-Slots protocol. Alice sends pulses of duration $T$ with chosen delay 0 or $T/2$. Bob measures the photon detection time. The time slots 1 and 3 are non ambiguous and allow for delay determination. The detections in time slot 2 are ambiguous and force Eve to introduce errors.}
\label{ThreeTimeSlot}
\end{figure}

In view of practical applications we have proposed a simple protocol based on time coding that makes use of coherent single photon pulses with square profile and duration $T$~\cite{Debuisschert2004,Boucher2005}. The key is encoded in the photon time-detection. The use of coherent superposition of states keeps Eve from eavesdropping the key without being noticed. This technique is attractive because it is one-way and it allows a simple implementation based on state-of-the-art optical components. A typical implementation has been proposed in~\cite{Debuisschert2004} and is depicted on figure~(\ref{TypicalScheme}).
We have considered several protocols based on such a principle. In the first one, depicted on figure~(\ref{ThreeTimeSlot}), we have considered three time-slots 1, 2 and 3 of duration $T/2$~\cite{Debuisschert2004}.
We call this protocol Three Time-Slots protocol (3TS).

Alice sends, at random and with equal probabilities, two kinds of pulses, encoding the bits 0 and 1. Bit 0 spans time-slots 1 and 2, bit 1 spans time-slots 2 and 3. Due to the non-orthogonality of the states encoding bit 0 and bit 1, it is impossible for the eavesdropper to preserve the coherence without introducing errors in the raw key.

Other protocols have been proposed based on related principles. The Differential-Phase-Shift (DPS) protocol sends temporal sequences of coherent pulses and encodes the key in the phase relation between successive pulses~\cite{Inoue2002}. The Coherent-One-Way (COW) protocol~\cite{Stucki2005,gisin2004} encodes the bits on pairs of adjacent time slots. In addition, it introduces superpositions of states spanning two time-slots. The COW protocol takes into account coherences between successive pulses, whereas in our protocol, we have considered internal coherences of the pulses. The exploitation of inter-pulses coherence aims at protecting against Photon-Number-Splitting (PNS) attacks~\cite{Brassard2000,gisin2004}. A complete proof of security is still to be performed on the COW protocol, although progress has been made~\cite{Branciard2008}.

\begin{figure}
\vspace*{13pt}
\includegraphics[width=0.6\columnwidth]{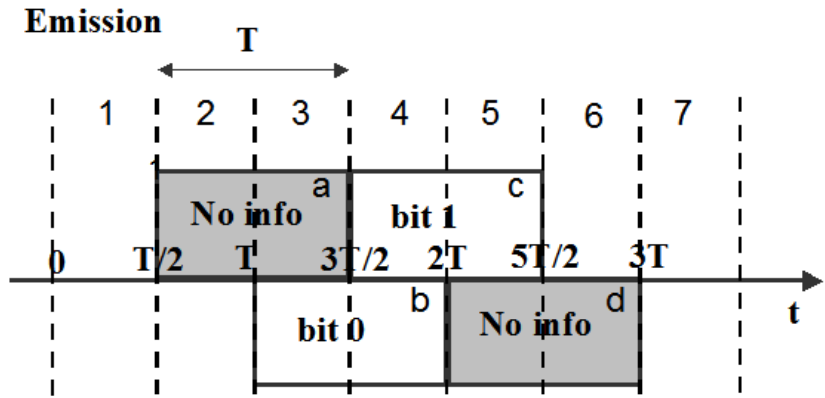}
\vspace*{13pt}
\caption{Principle of the four states protocol. Alice sends pulses of duration $T$ with chosen delays 0, $T/2$, $T$ or $3T/2$. Pulses (a) and (d) carry no information. Pulses (b) and (c) encode bit 0 and bit 1 respectively. Bob measures the photon detection time. He keeps only the results corresponding to time slot 3 and time slot 5. The results are ambiguous, which prevents Eve from exploiting the losses of the line.}
\label{FourStatesProtocol}
\end{figure}

In~\cite{Debuisschert2004}, we had seen that it was possible to eavesdrop the channel as soon as its losses exceed 50 \%. By doing this, Eve modifies the channel transmission for each time-slot and suppresses all the detection events corresponding to an ambiguous result (time-slot 2 in figure~(\ref{ThreeTimeSlot})). To remedy this drawback we had proposed a four states protocol where the two additional pulses carry no information and where successive pulses have a $T/2$ overlap~\cite{Debuisschert2004} (figure~\ref{FourStatesProtocol}). Recent related proposals also make use of additional states to make the B92 robust against losses~\cite{lucamarini2009robust}. It is in fact possible to keep the previous protocol with only three time-slots, if additional pulses of duration $T/2$ are combined at random with the pulses representing the bits. Measuring the number of such pulses received for each time-slot allows Alice and Bob to monitor the channel transmission for each time-slot and to ensure that it is not modified by Eve, which prevents attacks such as the previous one.

\begin{figure}
\vspace*{13pt}
\includegraphics[width=0.6\columnwidth]{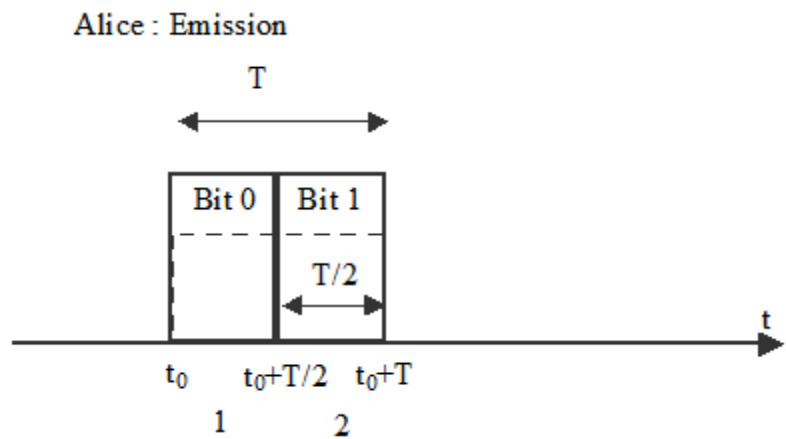}
\vspace*{13pt}
\caption{Principle of the Two Time-Slots protocol. Alice encodes the bits in successive non-overlapping single-photon pulses with duration $T/2$ (solid line). In addition, she sends at random single-photon pulses spanning the time-slots 1 and 2 allowing a coherence measurement aiming at detecting eavesdropping (dashed line).}
\label{TwoTimeSlot}
\end{figure}

In our previous works, we had considered only the simple case of intercept-resend attacks~\cite{Debuisschert2004,Boucher2005}. Our purpose is to generalize the study to optimal collective attacks where we consider the most general unitary transform allowed by quantum mechanics in order to upper bound the information that Eve can get on the key shared between Alice and Bob. The security of the key can be guaranteed when the mutual information between Alice and Bob $I_{AB}$ is greater than the mutual Holevo quantity between Alice and Eve $\chi_{AE}$~\cite{Csiszar1978,Devetak2005}.
In the present paper, we will first model the 3TS protocol where single-photon pulses spanning the time-slots (1,2) and (2,3) represent the bits 0 and 1 respectively.
From the quantum model describing this protocol, we deduce a simplification that allows to consider a simpler protocol. The bits are then encoded on two adjacent but non overlapping pulses (time-slots 1 and 2). Pulses spanning time-slots 1 and 2 are combined at random with the previous pulses. They allow to check that the coherence is not affected by eavesdropping.
We call this protocol the Two Time-Slots Protocol (2TS) as depicted in figure~(\ref{TwoTimeSlot}).
It has many similarities with BB84, and similar protocols have already been proposed~\cite{Branciard2007a,fung2006security}. The states corresponding to the H-V basis is represented by the time-slots 1 and 2. Only one state of the non-orthogonal basis is used which is the pulse spanning time slots 1 and 2.
We model Eve's attack introducing the most general unitary transform that could allow her to extract some information on the key. We perform the complete analysis of this protocol taking into account the imperfections of the system. We thus calculate the Holevo quantity between Alice and Eve, and the Shannon information between Alice and Bob as a function of the quantum bit error rate (QBER) for several values of the interferometer visibility. When the visibility of the interferometer is perfect, the expression of the Holevo quantity between Alice and Eve is identical to that obtained in the case of the BB84 protocol. The 2TS protocol can thus be viewed as a time-coding version of the BB84 protocol.

\begin{figure}
\vspace*{13pt}
\includegraphics[width=0.6\columnwidth]{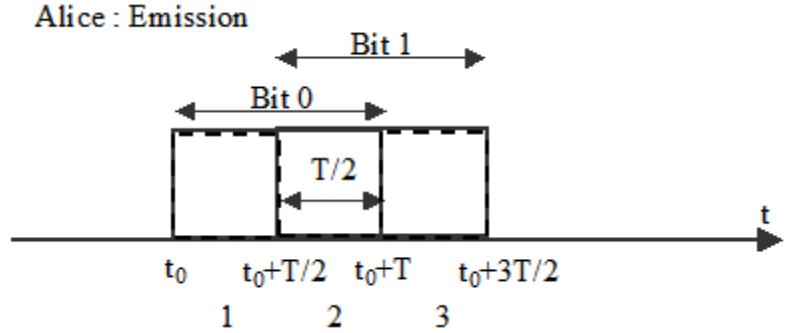}
\vspace*{13pt}
\caption{Principle of the Completed Three Time-Slots protocol. Alice encodes the bits in overlapping single-photon pulses spanning the time-slots 1-2 and 2-3 respectively (solid line). In addition, she sends at random single-photon pulses spanning the time-slots 1 and 3 allowing a coherence measurement aiming at detecting eavesdropping (dashed line).}
\label{CompletedThreeTimeSlot}
\end{figure}

The security analysis of the 2TS protocol is then used as an intermediate step to analyse the security of the 3TS protocol.
Although less performant than the 2TS protocol, this protocol is secure up to a QBER of $5\%$.
The main advantage of the 3TS protocol is that the pulses containing the information are similar up to a time shift. Whereas the 2TS protocol uses pulses of different length. This is an advantage in simplicity at the price of a reduction of the security.
The 3TS protocol can be improved if the pulses carrying the bits are completed with additional pulses featuring a coherent superposition between time-slots 1 and 3. We call this improved version Completed Three Time-Slots protocol (C3TS). It is depicted on figure~(\ref{CompletedThreeTimeSlot}). The results are then the same as for the 2TS protocol.

After having considered single-photon pulses in the previous cases, we then consider the case where Alice sends faint pulses instead of single-photons. In that case, the protocols are sensitive to photon number splitting attacks (PNS) where Eve keeps only the pulses where more than one photon is present, which allows her to get a perfect copy of the key. A counter-measure to such attack consists in introducing decoy states where the average value of the photon number in the pulses can take several values according to a given proportion~\cite{Hwang2003,Lo2005prl,Wang2005}. This keeps Eve from changing the number of photons in the pulses without being noticed. Using an approach closely related to that of~\cite{Gottesman2004,Lo2005prl,Ma2005pra,Yuan2007}, we calculate the rate as a function of the distance and compare it to that obtained in the case of single-photon pulses. As expected, we obtain a linear dependence with the channel attenuation in the case of single-photon pulses or decoy states, whereas it is quadratic without the use of decoy states. 

\section{Quantum formalism}
\label{sec:Quantum formalism}

The protocols introduced in the previous part can be modeled using a quantum formalism for the time-slots used to describe the pulses.
Although the time scale of Bob can be divided in an infinity of time-slots, a natural limit is given by the period between two successive pulses.
This period is divided in successive time-slots of duration $T/2$ numbered from 1 to $N$, which gives rise to a $N$ dimensions Hilbert space. To each time-slot corresponds a basis state $\ket{i}_B$ ($i$=1 to $N$) in Bob's Hilbert space.

We start with the 3TS protocol, where Bob's Hilbert space is three-dimensional with basis states noted $\ket{1}_B$, $\ket{2}_B$ et $\ket{3}_B$.
The pulses chosen by Alice to encode the bits are represented by the states $\ket{\beta_1}$ and $\ket{\beta_2}$. Since they have an overlap, they correspond to non-orthogonal states in Bob's space :
\begin{eqnarray}\label{beta1beta2}
  \ket{\beta_1}=\frac{\sqrt{2}}{2}(\ket{1}_B+\ket{2}_B)\\
  \ket{\beta_2}=\frac{\sqrt{2}}{2}(\ket{2}_B+\ket{3}_B)
\end{eqnarray}
Here we consider that Alice sends perfect pulses with no component of $\ket{\beta_1}$ on $\ket{3}_B$, and no component of $\ket{\beta_2}$ on $\ket{1}_B$.
The pulses are sent at random by Alice with identical probabilities equal to $\frac{1}{2}$. The state received by Bob is a statistical mixture given by :
\begin{equation}
  \rho_B=\frac{1}{2}\ket{\beta_1}\bra{\beta_1}+\frac{1}{2}\ket{\beta_2}\bra{\beta_2}.
\end{equation}
This description corresponds to the ``prepare and measure" description of the protocol. It describes the practical
way of preparing the pulses and sending them to Bob. An equivalent approach is called the ``virtual entanglement" description~\cite{Shor2000,grosshans:qic}. Since there is no eavesdropping at this stage, Alice and Bob are isolated. We can introduce a Hilbert space for Alice and describe the whole system by a pure state in the joint Hilbert space of Alice and Bob. Alice's Hilbert space is described by a two orthogonal states basis $\ket{1}_A$ and $\ket{2}_A$. These states can be associated with the bit 0 and the bit 1 respectively.
Before sifting, the state describing the system is :
\begin{equation}\label{psiAB}
  \ket{\psi_1}_{AB}=\left(\frac{\sqrt{2}}{2}\ket{1}_A\ket{\beta_1}
  +\frac{\sqrt{2}}{2}\ket{2}_A\ket{\beta_2}\right)\otimes\ket{0}_E
\end{equation}
The tensor product with the state $\ket{0}_E$ indicates that Eve does not interact with Alice and Bob. The partial trace of $_{AB}\ket{\psi_1}\bra{\psi_1}_{AB}$ over Alice results in the density matrix 
$\rho_B$. $\ket{\psi_1}_{AB}$ is thus a purification of $\rho_B$.

Eq.~(\ref{psiAB}) allows a detailed description of the protocol. We first evaluate the visibility of the interferometer, which allows to check the coherence of the pulses received by Bob. According to fig.~(\ref{TypicalScheme}), the photons detected at the output of the interferometer do not participate to the final secret key. Therefore, Bob can inform Alice when he has detected such a photon, and Alice can reveal which kind of pulse she has sent.
For example, we consider that Alice has sent a state $\ket{\beta_1}$. An additional two dimensional Hilbert space is introduced in order to describe the two arms of the interferometer with basis states $\ket{+}_I$ and $\ket{-}_I$. The incoming state is thus $\ket{in}_I=\frac{\sqrt{2}}{2}(\ket{1}_B+\ket{2}_B)\ket{+}_I$. After the first beamsplitter, the system is described with a superposition of the two interferometer states. Then, the state of one arm experiences the transform $\ket{i}_B\rightarrow\ket{i+1}_B$ in order to take into account the $T/2$ time difference between the two arms. Finally, introducing a possible phase-shift between the two arms, one ends up with the expression of the state at the output of the interferometer :
\begin{equation}\label{interfero}
  \ket{out}_I=
 \frac{\sqrt{2}}{4}\left[\left(\ket{1}_B+(1+e^{i\phi})\ket{2}_B+e^{i\phi}\ket{3}_B\right)\ket{+}_I
  +\left(\ket{1}_B+\left(1-e^{i\phi}\right)\ket{2}_B-e^{i\phi}\ket{3}_B\right)\ket{-}_I\right]
\end{equation}

From such an expression, it appears that, knowing that Alice has sent a state $\ket{\beta_1}$, Bob needs to keep only the detection events in time-slot 2 in order to measure the coherence of the pulse. In that case, a visibility of 1 is expected if no eavesdropping occurs. If a state $\ket{\beta_2}$ is incoming, a similar expression will be obtained with all the states $\ket{i}_B$ replaced with $\ket{i+1}_B$ in Eq.~(\ref{interfero}). Thus, a visibility of 1 can be obtained when keeping only the detection events in time-slot 3. Bob informs Alice each time he has detected a photon in time-slots 2 or 3 at the output of the interferometer. Alice then reveals whether she has sent a state $\ket{\beta_1}$ or a state $\ket{\beta_2}$. Bob can then calculate the visibility of the interferometer which is equal to 1 in case no eavesdropping occurs.

The state $\ket{\psi_1}_{AB}$ can be equivalently written in Bob's basis :
\begin{equation}\label{psiAB3}
  \ket{\psi_1}_{AB}=\left(\frac{1}{2}\ket{1}_A\ket{1}_B+\frac{1}{2}(\ket{1}_A
  +\ket{2}_A)\ket{2}_B+\frac{1}{2}\ket{2}_A\ket{3}_B\right)\otimes\ket{0}_E
\end{equation}

Using Eq.~(\ref{psiAB}), one obtains that the time slots with non zero probability detection are 1 and 3 (probability 1/4) and 2 (probability 1/2). When detecting the pulses sent by Alice, Bob ensures that these probability detections are respected.
The ket $\ket{2}_B$ is correlated to a superposition of $\ket{1}_A$ and $\ket{2}_A$, with equal weight. This state is thus ambiguous in the sense that it is impossible for Bob to know if Alice has sent a bit 0 or a bit 1. The projection of $\ket{\psi_1}_{AB}$ to the subspace ($\ket{1}_B$,$\ket{3}_B$) is given by
\begin{equation}\label{psiABsift2}
  \ket{\psi_{1-3}}_{AB}=\left(\frac{\sqrt{2}}{2}\ket{1}_A\ket{1}_B+\frac{\sqrt{2}}{2}\ket{2}_A\ket{3}_B\right)\otimes\ket{0}_E
\end{equation}
This state is maximally entangled between Alice and Bob. It denotes a complete correlation between the bit chosen by Alice and the detection in Bob's non-ambiguous time-slots 1 and 3.

The expression of $\ket{\psi_1}_{AB}$ given by Eq.~(\ref{psiAB3}) can be interpreted as a new protocol where Bob sends at random the states $\ket{1}_A$, $\ket{2}_A$ or $\frac{\ket{1}_A+\ket{2}_A}{\sqrt{2}}$ with respective probabilities of 1/4, 1/4 and 1/2. $\ket{1}_A$ and $\ket{2}_A$ are used to encode the key. The state $\frac{\ket{1}_A+\ket{2}_A}{\sqrt{2}}$ is used to ensure that the eavesdropper does not break the coherence between $\ket{1}_A$ and $\ket{2}_A$. It imposes a constraint on eavesdropping which ensures the security of the protocol.
After the pulses have been sent, sifting occurs, and Bob informs Alice when he has sent states $\frac{\ket{1}_A+\ket{2}_A}{\sqrt{2}}$. To establish the key, Alice and Bob keep only the states corresponding to Bob sending either $\ket{1}_A$ or $\ket{2}_A$.
The state representing the system after sifting is given by Eq.~(\ref{psiABsift2}). Eq.~(\ref{psiAB3}) thus describes a protocol dual from that described by Eq.~(\ref{psiAB}). Finally, since Eq.~(\ref{psiABsift2}) is symmetrical, the roles of Alice and Bob can be interchanged and one can assume that Alice sends the pulses and Bob detects the photons. We then have two protocols where Alice sends the pulses and Bob detects the photons. The one described by Eq.~(\ref{psiAB}) uses a three-dimensional Hilbert space for Bob and
corresponds to the 3TS protocol.
It is the initially proposed time-coding protocol. The second protocol described by Eq.~(\ref{psiAB3}) uses only a two-dimensional Hilbert space for Bob and
corresponds to the 2TS protocol.
The security analysis of this latter is simpler and we will first analyse it. We will then use the main lines of the security analysis of the 2TS protocol to analyse the security of the 3TS protocol.

\section{Security analysis of the Two Time-Slots protocol}\label{partTwoBasisState}

The state describing the joint Hilbert space of Alice and Bob, when the pulses are sent by Alice and no eavesdropping occurs, can be written :

\begin{equation}\label{psiAB4}
  \ket{\psi_2}_{AB}=\left(\frac{1}{2}\ket{1}_B\ket{1}_A+\frac{1}{2}(\ket{1}_B
  +\ket{2}_B)\ket{3}_A+\frac{1}{2}\ket{2}_B\ket{2}_A\right)\otimes\ket{0}_E
\end{equation}
As compared to Eq.~(\ref{psiAB3}), we have reversed the roles of Alice and Bob and we have interchanged the kets $\ket{2}_A$ and $\ket{3}_A$ to have more symmetrical notations. Thus $\ket{1}_A$ and $\ket{2}_A$ describe the pulses carrying the bits and $\ket{3}_A$ describes the pulses carrying the coherence information.

We then consider the general attack allowing Eve to be entangled with the states sent by Alice to Bob~\cite{Bruss2002,gisin:rmp,PhysRevA.56.1163}.
The two states of Bob are transformed according to the following relations.
\begin{eqnarray}  \label{transfo2etats}
    \ket{1}_B\ket{0}_E\rightarrow\ket{1}_E=\sqrt{F_1}\,\ket{11}\,\ket{1}_B+\sqrt{Q_1}\,\ket{12}\,\ket{2}_B\label{transfo2etats1}\\
    \ket{2}_B\ket{0}_E\rightarrow\ket{2}_E=\sqrt{F_2}\,\ket{22}\,\ket{2}_B+\sqrt{Q_2}\,\ket{21}\,\ket{1}_B\label{transfo2etats2}
\end{eqnarray}

In this transform, Eve's Hilbert space is described by the states $\ket{ij}$ where $i$ and $j$ refer to the initial state and to the final state of Bob, respectively. The normalisation condition implies $F_1+Q_1=F_2+Q_2=1$. We consider that the attack of Eve can be asymmetric in the case where $F_1\neq F_2$. Anyhow, we will show in the following that the symmetric attack is optimal.
The joint Hilbert space of Alice, Bob and Eve is a closed system. Therefore, the transform must be unitary,
which implies the relation :
\begin{equation}\label{ortho22}
  _E\braket{1}{2}_E=\sqrt{F_1 Q_2}\braket{11}{21}+\sqrt{F_2 Q_1}\braket{22}{12}=0;
\end{equation}
The state resulting from Eve's attack is given by :
\begin{equation}\label{psiABE2state}
  \ket{\psi}_{ABE}=\frac{1}{2}\ket{1}_E\ket{1}_A+\frac{1}{2}(\ket{1}_E
  +\ket{2}_E)\ket{3}_A+\frac{1}{2}\ket{2}_E\ket{2}_A
\end{equation}
Classical communication allows Alice to inform Bob whether she has sent a state carrying the bit information (subspace ($\ket{1_A}$,$\ket{2_A}$)) or a state carrying the coherence information (subspace ($\ket{3_A}$)).
In order to consider the states carrying the bit information, Bob projects the state $\ket{\psi}_{ABE}$ on the subspace ($\ket{1_A}$,$\ket{2_A}$) given by :
\begin{equation}\label{psiABE2statesift}
  \ket{\psi_{1-2}}_{ABE}=\frac{\sqrt{2}}{2}[\sqrt{F_1}\ket{11}\ket{1}_A\ket{1}_B+\sqrt{F_2}\ket{22}\ket{2}_A\ket{2_B}
  +\sqrt{Q_1}\ket{12}\ket{1}_A\ket{2}_B+\sqrt{Q_2}\ket{21}\ket{2}_A\ket{1}_B]
\end{equation}

The state described by Eq.~(\ref{psiABE2statesift}) is an entangled pure state between Alice and Bob on one hand and Eve on the other hand. The entanglement witnesses the correlations between Eve's measurement and Alice and Bob's measurement. The corresponding information can be quantified by the entropy of entanglement~\cite{haroche2006exploring}.
Maximizing this quantity, Eve maximizes the knowledge she has on the key. From the state $\ket{\psi_{1-2}}_{ABE}$, one can deduce the density matrix of the complete system $\rho_{{1-2}_{ABE}}=\ket{\psi_{1-2}}_{ABE}\,_{ABE}\bra{\psi_{1-2}}$. The density matrix of the reduced system is defined by $\rho_{AB}=\mathrm{tr}_E(\rho_{{1-2}_{ABE}})$ for Alice and Bob and $\rho_{E}=\mathrm{tr}_{AB}(\rho_{{1-2}_{ABE}})$ for Eve.
The entropy of entanglement is defined by :
\begin{equation}\label{SrhoAB=SrhoE}
  S_{ent}=S(\rho_{AB})=S(\rho_{E})
\end{equation}
where $S(\rho)=-\mathrm{tr}(\rho \log_2(\rho))$ is the Von Neuman entropy of $\rho$.
In order to maximize her information, Eve has to maximize $S(\rho_{AB})$.

$\rho_{AB}$ is a $4\times4$ matrix in the basis ($\ket{1}_A\ket{1}_B$, $\ket{2}_A\ket{2}_B$, $\ket{1}_A\ket{2}_B$, $\ket{2}_A\ket{1}_B$).
$\ket{1}_A\ket{1}_B$ and $\ket{2}_A\ket{2}_B$ form a basis for the subspace corresponding to Alice and Bob getting identical results. $\ket{1}_A\ket{2}_B$ and $\ket{2}_A\ket{1}_B)$ form a basis for the subspace corresponding to Alice and Bob getting different results.

Let us apply the projective measurement on those two subspaces to $\rho_{AB}$. The resulting density matrix $\rho_{AB}'$ is given by :

\begin{equation}\label{rhoAB'}
\rho_{AB}'=\frac{1}{2}\left(\begin{array}{cccc}
F_1 & \sqrt{F_1 F_2}\braket{11}{22} & 0 & 0\\
\sqrt{F_1 F_2}\braket{22}{11} & F_2 & 0 & 0\\
0 & 0 & Q_1 & \sqrt{Q_1 Q_2}\braket{12}{21}\\
0 & 0 & \sqrt{Q_1 Q_2}\braket{21}{12} & Q_2
\end{array}\right)	
\end{equation}

As a result of a projective measurement, we have
\begin{equation}
  S(\rho_{AB}')\geq S(\rho_{AB})
\end{equation}
$S(\rho_{AB})$ is maximized when $\rho_{AB}$ is chosen identical to $\rho_{AB}'$. This results from the loss of information when the non-diagonal blocks are taken equal to zero in $\rho_{AB}$.
Each term of the non-diagonal blocks of $\rho_{AB}$ depends only on one of the following scalar products : $\braket{11}{12}$, $\braket{11}{21}$, $\braket{22}{12}$ and $\braket{22}{21}$. Therefore, an optimal choice for Eve in order to maximize her information on the system is to choose those four scalar products equal to zero.

As a result, the unitarity condition Eq.~(\ref{ortho22}) is automatically fulfilled. In addition, the expressions of $\ket{1}_E$ and $\ket{2}_E$ are Schmidt decompositions for which Eve and Bob are maximally entangled. The attack performed by Eve introduces some errors.
Since $\braket{11}{12}=0$ and $\braket{22}{21}=0$ the corresponding noise is incoherent and cannot be distinguished from any real noise coming from the imperfections of the experiment.

The choice of the scalar products above is a first step for Eve to maximize her information on Alice and Bob's system. A further step would be to choose the two remaining scalar products equal to zero as well.
Anyhow, the states of Eve would all be orthogonal one with another, and this would result in the impossibility to measure any interference at Bob's interferometer, which is a prerequisite for the protocol to be useful. Thus $\braket{11}{22}$ and $\braket{12}{21}$ are left as free parameters the value of which can be optimized by Eve.

According to Eq.~(\ref{SrhoAB=SrhoE}) and considering the optimal choice $\rho_{AB}=\rho_{AB}'$, $S\left(\rho_{E}\right)$ can be calculated from the eigenvalues of Eq.~(\ref{rhoAB'}) using the definition $S\left(\rho_{E}\right)=\sum_{i=1}^4{-\gamma_i \log_2(\gamma_i)}$. Defining $F=\frac{1}{2}(F_1+F_2)$, $Q=\frac{1}{2}(Q_1+Q_2)$ and $dQ=\frac{1}{2}(Q_1-Q_2)$, the expressions of the eigenvalues are :
\begin{eqnarray}
    \gamma_1=\frac{1}{2}\left(F+\sqrt{\left(1-{{
\braket{11}{22}}}^{2}\right){{ dQ}}^{2}+{{ \braket{11}{22}}}^{2}{F}^{2}}\right)\\
  \gamma_2=\frac{1}{2}\left(F-\sqrt{\left(1-{{
\braket{11}{22}}}^{2}\right){{ dQ}}^{2}+{{ \braket{11}{22}}}^{2}{F}^{2}}\right)\\
\gamma_3=\frac{1}{2}\left(Q+\sqrt{\left(1-{{
\braket{12}{21}}}^{2}\right){{ dQ}}^{2}+{{ \braket{12}{21}}}^{2}{Q}^{2}}\right)\\
\gamma_4=\frac{1}{2}\left(Q-\sqrt{\left(1-{{
\braket{12}{21}}}^{2}\right){{ dQ}}^{2}+{{ \braket{12}{21}}}^{2}{Q}^{2}}\right)
\end{eqnarray}
where $-\frac{1}{2}\leq dQ\leq\frac{1}{2}$ and $|dQ|\leq Q\leq 1-|dQ|$.
Studying the variations of $S\left(\rho_{E}\right)$ with $dQ$ shows that it is maximal when $dQ=0$, due to dependence of  $S\left(\rho_{E}\right)$ with the square of $dQ$. Therefore, Eve can maximize her entanglement with Alice and Bob choosing $F_1=F_2=F$ and $Q_1=Q_2=Q$.

The information available for Eve is upper bounded by the Holevo quantity defined by :
\begin{equation}\label{khiAE}
\chi=S\left(\rho_{E}\right)-\sum_{i}p_iS\left(\rho_{E}^{i}\right)
\end{equation}
In the case where $\chi_{AE}$ is calculated, the matrix $\rho_{E}^{i}$ is the density matrix of Eve when one knows which state has been sent by Alice.
From Eq.~(\ref{psiABE2state}), those two states are $\ket{1}_E$ and $\ket{2}_E$ and one gets :
\begin{equation}
  \rho_{E}^{1}=\left(\begin{array}{cc}
F_1 & 0\\
0 & Q_1
\end{array}\right),
  \rho_{E}^{2}=\left(\begin{array}{cc}
Q_2 & 0\\
0 & F_2
\end{array}\right)
\end{equation}
Since those states are equiprobable,we have $p_1=p_2=\frac{1}{2}$, and the Holevo quantity becomes
\begin{equation}\label{chiAE}
  \chi_{AE}=S\left(\rho_{E}\right)-\frac{1}{2}h(Q_1)-\frac{1}{2}h(Q_2)
\end{equation}
where $h(Q)=-Q\log_2(Q)-(1-Q)\log_2(1-Q)$.

The available secret bit per sifted pulse is given by
\begin{eqnarray}\label{DeltaI}
\Delta I=I_{AB}-\chi_{AE}
\end{eqnarray}
$I_{AB}$ is the mutual information between Alice and Bob. In order to compute $I_{AB}$, we have to take into account that the error rates are different, depending on which state $\ket{1}_A$ or $\ket{2}_A$ is sent by Alice. Alice and Bob can measure those two errors rates revealing a fraction of the pulses that are sent. Since those pulses are equiprobable, the mutal information is the average of the mutual information corresponding to each state, and we obtain :
\begin{equation}\label{IAB}
  I_{AB}=1-\frac{1}{2}\left(h(Q_1)+h(Q_2)\right)
\end{equation}
Combining Eq.~(\ref{chiAE}) and Eq.~(\ref{IAB}), we obtain the expression of the secret key rate :
\begin{equation}
  \Delta I=1-S\left(\rho_{E}\right)
\end{equation}
This expression shows that the secret bit rate is directly related to the entropy of entanglement. The goal of Eve is to minimize the secret bit rate and thus to maximize $S\left(\rho_{E}\right)$. According to the previous derivation, we obtain that this occurs when $dQ=0$, thus showing that the symmetric attack is optimal. In the following of the paper, we will thus consider that Eve's attacks are symmetric.

In the next step, we optimize $S\left(\rho_{E}\right)$ for a given value of the fringe visibility measured at Bob's interferometer $V_{12}$, which is measured when Alice sends a superposition state of the form $\frac{1}{\sqrt{2}}(\ket{1}_B+\ket{2}_B)$, and Bob takes into account the photons detected in time-slot 2. Due to Eve's action, $V_{12}$ can be smaller than 1. Its expression is given by :
\begin{equation}
  V_{12}=F\braket{22}{11}+Q\braket{12}{21}
\end{equation}
We find that $S\left(\rho_{E}\right)$ is maximum when $\braket{12}{21}=\braket{11}{22}=V_{12}$. $S_{max}\left(\rho_{E}\right)$ is thus given by:
\begin{equation}
  S_{max}\left(\rho_{E}\right)=\sum_{i=1}^4 -\lambda_i \log_2(\lambda_i)
\end{equation}
with :
\begin{eqnarray}
  \lambda_1&=&\frac{1}{2}Q(1+V_{12})\label{lambda1}\\
  \lambda_2&=&\frac{1}{2}Q(1-V_{12})\\
  \lambda_3&=&\frac{1}{2}F(1+V_{12})\\
  \lambda_4&=&\frac{1}{2}F(1-V_{12})\label{lambda4}
\end{eqnarray}

We can then model the imperfections of the real channel between Alice and Bob. Alice sends a state represented by a density matrix $\rho_A$. The channel is characterized by a global transmission $\eta$. Assuming that the channel is symmetric and that the induced modification is independent from the initial state, we obtain a resulting state at Bob that has the following expression.
\begin{equation}\label{rhoB}
  \rho_B=\eta\rho_A+\frac{1-\eta}{2}I_2
\end{equation}
Where $I_2$ is the $2\times2$ identity matrix. In case where the initial state is $\ket{1}_B$ or $\ket{2}_B$, one obtains that the fidelity is $\frac{1}{2}(1+\eta)$ and that the error rate is $\frac{1}{2}(1-\eta)$. In case where the initial state is a partially coherent superposition of states characterized by a visibility $V_A$, the resulting visibility at Bob is $V_B=\eta V_A$. The lack of visibility at Bob occurs from two origins. The first one, represented by $V_A$, is due to the intrinsic imperfections of the apparatus or a possible lack of coherence of the source. The second one, represented by $\eta$, is due to the imperfect transmission channel.

\begin{figure}
\vspace*{13pt}
\includegraphics[width=0.6\columnwidth]{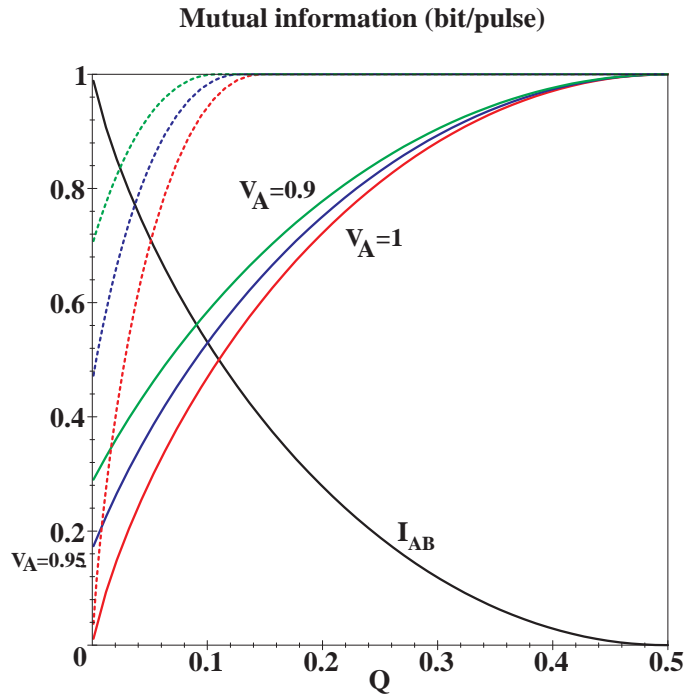}
\vspace*{13pt}
\caption{Mutual informations as a function of the quantum bit error rate Q. The curves represent the Holevo quantity between Alice and Eve for visibilities $V_A$ equal to 0.9, 0.95 and 1. The solid lines represent Eve's information in the case of the Two Time-Slots protocol or the Completed Three Time-Slots protocol. For a visibility of 1, it crosses the curve representing the Shannon information between Alice and Bob, $I_{AB}$, for a quantum bit error rate of 0.11. The dashed lines represent Eve's information in the case of the original Three Time-Slots protocol for the same values of the visibility. This procotol is less performant than the previous ones as shown for the maximal QBER and the available information per pulse that are smaller. Anyhow, security is still possible up to values of QBER equal to $5 \%$ when $V_A=1$.}
\label{figureIABIAE_2Etats}
\end{figure}

Eve substitutes her perfect measurement apparatus, represented by the unitary transform (Eq.~(\ref{transfo2etats1}) and Eq.~(\ref{transfo2etats2})), to the imperfect channel between Alice and Bob. This is equivalent to a channel with transmission $\eta=F-Q$. The visibility measured at Bob is thus $V_B=(F-Q)V_A$.
It is assumed that Eve exploits all imperfections of the set-up, and thus she can set $V_{12}=V_B=(F-Q)V_A$. The values involved in $S\left(\rho_{E}\right)$ can be expressed by :
\begin{eqnarray}\label{eigenvals}
  \lambda_1&=&\frac{1}{2}Q(1+(F-Q)V_A)\label{eigenvals1}\\
  \lambda_2&=&\frac{1}{2}Q(1-(F-Q)V_A)\label{eigenvals2}\\
  \lambda_3&=&\frac{1}{2}F(1+(F-Q)V_A)\label{eigenvals3}\\
  \lambda_4&=&\frac{1}{2}F(1-(F-Q)V_A)\label{eigenvals4}
\end{eqnarray}
From those expressions, we can compute the information curves represented on figure (\ref{figureIABIAE_2Etats}) for several values of $V_A$ as a function of $Q$. In the particular case of a perfect interferometer $(V_A=1)$, the expression of $\chi_{AE}$ simplifies to $\chi_{AE_1}=h(Q)$, and the secret key rate becomes $1-2h(Q)$. It coincides with the bound given by GLLP~\cite{Gottesman2004} for the BB84 protocol when Eve launches a basis independent attack, with a maximum QBER of 0.11.

\section{Security analysis of the Three Time-Slots protocol}

After analyzing the security of the two time-slots protocol (2TS), we can go back to the original three time-slots protocol (3TS). The state describing the whole system, when no eavesdropping occurs, is given by Eq.~(\ref{psiAB}). The state corresponding to the projection on the basis states $\ket{1}_B$ and $\ket{3}_B$ is given by Eq.~(\ref{psiABsift2}).
The key is built upon those two states, and therefore, the attack of Eve should involve only them and not the state $\ket{2}_B$ which does not play any role after sifting. As a consequence, the unitary transform describing the attack of Eve is similar to that used in the case of the 2TS protocol (Eq.~(\ref{transfo2etats1}) and Eq.~(\ref{transfo2etats2})), the role of $\ket{2}_B$ being played by $\ket{3}_B$. The state $\ket{2}_B$ is transformed in a way similar to that of $\ket{1}_B$ and $\ket{3}_B$. A fourth state $\ket{4}_B$ is introduced to preserve the unitarity of the transform but it does not play any role in the security evaluation. Doing this, we find the same fidelity for all three states after Eve's attack.
The unitary transform applied by Eve thus writes :
\begin{eqnarray}\label{transfo3etats}
    \ket{1}_B\ket{0}_E\rightarrow\ket{1}_E&=&\sqrt{F}\,\ket{11}\,\ket{1}_B+\sqrt{Q}\,\ket{13}\,\ket{3}_B\label{transfo3etats1}\\
    \ket{2}_B\ket{0}_E\rightarrow\ket{2}_E&=&\sqrt{F}\,\ket{22}\,\ket{2}_B+\sqrt{Q}\,\ket{24}\,\ket{4}_B\label{transfo3etats2}\\
    \ket{3}_B\ket{0}_E\rightarrow\ket{3}_E&=&\sqrt{F}\,\ket{33}\,\ket{3}_B+\sqrt{Q}\,\ket{31}\,\ket{1}_B\label{transfo3etats3}
\end{eqnarray}
In the case of the 2TS protocol, the security relies on a  coherence measurement corresponding to Alice sending a superposition state $\frac{1}{\sqrt{2}}(\ket{1}_B+\ket{2}_B)$. Similarly, 
in this case, Alice should send a superposition state $\frac{1}{\sqrt{2}}(\ket{1}_B+\ket{3}_B)$ between the two states involved in the key establishment. Then Bob should measure the coherence between those two states after transmission through the channel, in order to evaluate the security of the exchange. In that case, one would expect to derive the same security analysis and to obtain exactly the same results as derived in part $\ref{partTwoBasisState}$. In particular, the information of Eve is maximized when the relation $\braket{13}{31}=\braket{11}{33}=V_{13}$ is fulfilled. As compared to the original 3TS protocol, this implies for Alice to send additional superpositions of pulses spanning time-slots 1 and 3 and for Bob to measure their coherence with an interferometer having a time propagation difference of  $T$ between the two arms. We will denote this protocol Completed Three Time-Slots protocol (C3TS).

Similarly to the C3TS protocol, the original 3TS protocol makes use of $\ket{1}_B$ and $\ket{3}_B$ to carry the information. Therefore, the security analysis of the 3TS protocol is the same as that of the 2TS and C3TS protocols up to the expressions of the eigenvalues of $S(\rho_E)$ as a function of the visibility (Eq.~(\ref{lambda1}) to Eq.~(\ref{lambda4})). The difference between the 3TS protocol and the C3TS protocol lies in the way the visibility is measured in both cases. In the C3TS protocol, the visibility is measured between time-slots 1 and 3.
In the original 3TS protocol, the interferometer measures the coherence between time-slots 1 and 2 and between time slots 2 and 3 respectively, as described in section~\ref{sec:Quantum formalism}. If Bob receives a superposition of states $\ket{1}_B$ and $\ket{2}_B$ he keeps only the photons detected in time-slot 2. If he receives a superposition of states $\ket{2}_B$ and $\ket{3}_B$ he keeps only the photons detected in time-slot 3. Therefore he can measure a visibility of 1 if no eavesdropping occurs. Taking into account Eq.~(\ref{transfo3etats1}) to Eq.~(\ref{transfo3etats3}), we can then calculate the visibility in case eavesdropping occurs, and we get :
\begin{eqnarray}
  V_{12}=\braket{11}{22}\\
  V_{23}=\braket{22}{33}
\end{eqnarray}

We can then relate the parameters $F$ and $Q$ to the parameters of the unperfect channel $\eta$ and $V_A$ as we did in section \ref{partTwoBasisState} for the 2TS protocol. Since the key is established using a two states Hilbert space ($\ket{1}_B$, $\ket{3}_B$), we can derive the relations $F=\frac{1}{2}(1+\eta)$ and $Q=\frac{1}{2}(1-\eta)$ as we did previously, using Eq.~(\ref{rhoB}). Similarly, the visibility evaluation involves two dimension Hilbert spaces, either ($\ket{1}_B$, $\ket{2}_B$) or ($\ket{2}_B$, $\ket{3}_B$). We then get similar values for the visibilities measured at Bob : $V_{12}=V_{23}=\eta V_A$. Finally, we get the relation :
\begin{equation}
  \braket{11}{22}=\braket{22}{33}=(F-Q)V_A=\cos(\phi)
\end{equation}

Similarly to the case of the C3TS protocol, Eve should optimize the angle between states $\ket{11}$ and $\ket{33}$ in order to minimize the scalar product $\braket{11}{33}$ and thus maximise her information. The states $\ket{11}$, $\ket{22}$ and $\ket{33}$ form a three dimension space.  The scalar product $\braket{11}{33}$ is thus constrained by the previous relation since $\ket{11}$ and $\ket{33}$ both make an angle $\phi$ with $\ket{22}$.
As long as $\phi\leq\frac{\pi}{4}$, the minimal value of the coherence between time-slots 1 and 3 is thus obtained when $\ket{11}$, $\ket{22}$ and $\ket{33}$ are in the same plane, and its value is given by
\begin{equation}
  V_{13}=\braket{11}{33}=\cos(2\phi)
\end{equation}
A simple trigonometric expansion shows that $(F-Q)V_A$ should be replaced by $2(F-Q)^2V_A^2-1$ in Eq.~(\ref{eigenvals1}) to Eq.~(\ref{eigenvals4}). The eigenvalues used to calculate the secure key rate become :
\begin{eqnarray}\label{eigenvalsThreeState}
  \lambda_1&=&Q(F-Q)^2V_A^2\\
  \lambda_2&=&Q(1-(F-Q)^2V_A^2)\\
  \lambda_3&=&F(F-Q)^2V_A^2\\
  \lambda_4&=&F(1-(F-Q)^2V_A^2)
\end{eqnarray}
If $\phi\geq\frac{\pi}{4}$, $\ket{11}$ and $\ket{33}$ can be chosen orthogonal, and Eve has a complete information on the key.

This protocol is less performant than the 2TS protocol or the C3TS protocol (figure (\ref{figureIABIAE_2Etats})).
In particular, Eve can obtain a complete information on the key as soon as $2(F-Q)^2V_A^2-1=0$, which corresponds to $Q=0.15$ in the case where $V_A=1$ . For the two other protocols, this occurs when $Q=0.5$ independently of the value of $V_A$. In the case of a perfect visibility ($V_A=1$), the maximum QBER is $5\%$, whereas it is $11\%$ in the case of the 2TS protocol or the C3TS protocol.

\section{Evaluation of the secure distance}
\label{sec:Evaluation of the secure distance}

One of the main characteristics of a quantum key distribution link is its secure distance, i.e. the distance over which the mutual information $I_{AB}$ stays greater than $\chi_{AE}$~\cite{Csiszar1978}. Our security evaluation involves two parameters which are the fringe visibility $V_A$ and the QBER $Q$. In order to eavesdrop the key, Eve replaces the imperfect channel with transmission $\eta$ with a perfect one having a transmission equal to 1. We assume in addition that Eve can control all the errors of the set-up. She can thus perform an attack that introduces errors up to a maximum QBER of $Q$. The QBER is dependent on the channel parameters.
It results from the errors intrinsic to Alice, $Q_A$, and from the dark counts in Bob detectors. In the case of the 2TS protocol, an error occurs, for example, when Alice sends a state $\ket{1_B}$ and Bob detects $\ket{2_B}$. In the case of the 3TS protocol, an error occurs when Alice sends for example a state $\ket{\beta_1}$ and Bob detects a state $\ket{3_B}$. The probability to detect in the wrong time-slot comes either from the intrinsic errors of the pulse sent by Alice affected by the channel transmission, $\eta Q_A$, or from the dark-count probability per time-slot, $p_d$. The total probability detection results either from an incoming photon with probability $\eta$ or from the dark-count probability corresponding to the  two time slots involved in the protocol, $2 p_d$. Therefore, the overall quantum bit error rate is given by :
\begin{equation}\label{Qexp}
  Q=\frac{\eta Q_A+p_d}{\eta+2p_d}
\end{equation}

In particular, $Q$ increases as $\eta$ decreases. When the transmission is close to 1, the probability of a dark-count is totally negligible and the QBER is equal to the QBER intrinsic to Alice $Q_A$. When the transmission becomes very low, the QBER is dominated by the dark-counts and it tends to $1/2$. The dependence of the channel transmission with distance is given by $\eta=\eta_d\eta_L$ where $\eta_d$ is the quantum efficiency of the detector, and $\eta_L={10}^{-\frac{\alpha L}{10}}$ is the channel attenuation where $L$ is the distance in km and $\alpha$ is the attenuation in dB/km. We consider a fiber attenuation $\alpha$ = 0.2 dB/km, a superconducting single-photon detector~\cite{Goltsman2001} with a dark-count rate of 10 counts/sec, a quantum efficiency of 10 \% and a pulse duration of $T/2$~=~10~ns. This leads to a dark-count probability per pulse $ p_d=10^{-7}$. For both protocols and for $V_A$ equal to 1, 0.95 and 0.9 and $Q_A=0.02$, we deduce the secure distance from the maximum allowed QBER (see figure (\ref{figureIABIAE_2Etats})).
For the original 3TS protocol, we obtain 227, 213 and 182~km. For the 2TS protocol or for the C3TS protocol
we obtain 253, 250 and 247 km. For a given value of the visibility, the 2TS or the C3TS protocols feature a longer secure distance than the original 3TS protocol. This comes clearly form the mutual information curves shown on figure $\ref{figureIABIAE_2Etats}$. Anyhow the difference between both protocols becomes really important for low values of the visibility $V_A$.
This evaluation shows that the 2TS protocol or the C3TS protocol
are better than the original 3TS protocol in term of secure distance and that secure distances over 200 km are possible with the combination of time-coding protocols and low-noise SSPD.

\section{Photon number splitting attacks}
\label{sec:Photon number splitting attacks}

Up to now, we have considered single-photon states and the security analysis has been done within the framework of this hypothesis. Although the subject of intense experimental research, single-photon states are still difficult to produce. Most of the time, QKD devices make use of faint pulses simulating single-photon states when sufficiently attenuated. Here, we will analyse the security of the 2TS protocol within this framework.

\begin{figure}
\vspace*{13pt}
\includegraphics[width=0.6\columnwidth]{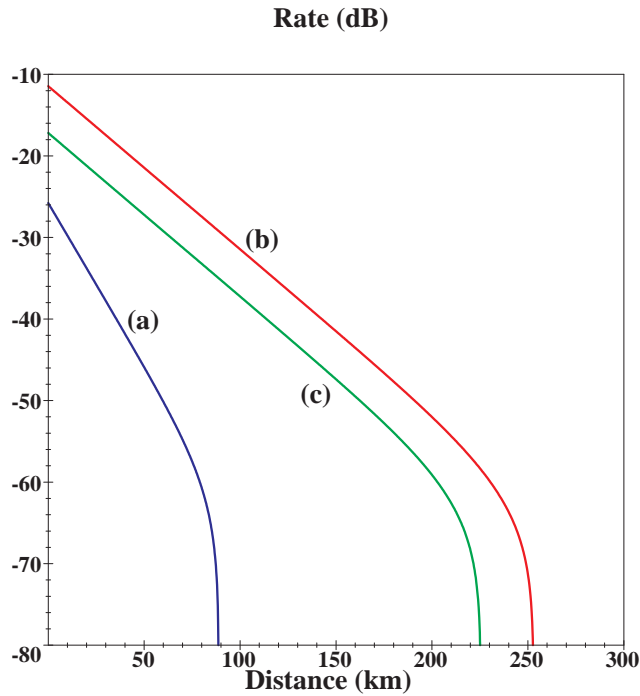}
\vspace*{13pt}
\caption{Decrease of the secure key rate (dB) as a function of the distance (km) for the 2TS protocol with visibility 1. Curve (a) is obtained with faint pulses without decoy states. The rate decreases as $\eta^2$ until it reaches the cut-off distance. Curve (b) is obtained with single photons, curve (c) is obtained combining faint pulses with decoy states. Both curves decrease as $\eta$ with the distance as long as $\eta\geq p_d$. When we consider a fiber attenuation of $\alpha$ = 0.2 dB/km, the cut-off distance is 90 km without decoy states, 250 km for the single photon protocol and 225 km for the decoy state protocol.}
\label{figureTauxDistance}
\end{figure}

The fact that more than one photon can be measured in a pulse makes possible the use of specific attacks on the key exchange. One of them, called \emph{Photon number splitting attack} (PNS)~\cite{Brassard2000}, explicitly exploits the number of photons in the pulse. The principle consists for Eve in measuring the number of photons in a given pulse and keeping only those with a number of photons greater than 1. Then, Eve keeps one photon and stores it in a quantum memory while she lets the remaining part of the pulse be transmitted to Bob. She waits until the end of the reconciliation process and the corresponding classical information exchange between Alice and Bob. She then performs the adequate measurement, in order to get some knowledge on the key.
The limit rate taking into account PNS attacks has been given by GLLP~\cite{Gottesman2004}. It can be written
\begin{equation}\label{tauxfaint}
R_{faint}\geq q G_\mu\{(1-\Delta)(1-h(Q_\mu/(1-\Delta)))-h(Q_\mu)\}
\end{equation}
$\Delta$ is the ratio between the number of multiphoton pulses measured at the output of Alice to the number of pulses actually detected by Bob. Therefore, $(1-\Delta)$ is a lower bound of the proportion of single photon pulses in the signal pulses. The first term in the parenthesis represents the mutual information $I_{AB}$ that can be obtained only from single photon pulses. The second term is the information available to the eavesdropper that has the same expression as in the case of single photon pulses. $q$ is a parameter specific from the protocol, $G_\mu$ and $Q_\mu$ are the gain and the error rate for the signal pulse, that can be evaluated from the channel characterization. In order to calculate each term in Eq.~(\ref{tauxfaint}), we start from the general expression of a two time-slot pulse with average photon number $\mu$. It is described by a product of coherent states
\begin{equation}\label{psiAgen}
\ket{\psi}_{Agen}=\ket{a_1\sqrt{\mu}}_1\ket{a_2\sqrt{\mu}}_2
\end{equation}
where $|a_1|^2+|a_2|^2=1$ and $|a_2|^2\ll|a_1|^2$ or $|a_1|^2\ll|a_2|^2$, depending on whether a bit 0 or a bit 1 is encoded.
Bob keeps only the events corresponding to one detection in a time-slot, and no detection in the other one. Therefore, the detection in time-slot 1 is described by the operator
${P}_1=({1}-\ket{0}_1\,_1\bra{0})\ket{0}_2\,_2\bra{0}$ and the detection in time-slot 2 by ${P}_2=({1}-\ket{0}_2\,_2\bra{0})\ket{0}_1\,_1\bra{0}$. Taking into account the dark-count probability $p_d$, we obtain the probability to have a detection in time-slot 1 (and none in time-slot 2) given by $p_1=(1-\exp(-|a_1|^2\eta\mu))\exp(-|a_2|^2\eta\mu)+p_d$. Similarly, for time-slot 2, we have $p_2=(1-\exp(-|a_2|^2\eta\mu))\exp(-|a_1|^2\eta\mu)+p_d$. At the output of the channel, the initial coherent~state $\ket{\sqrt{\mu}}$ is transformed to $\ket{\sqrt{\eta\mu}}$, due to the overall channel transmission. For small values of $\mu$, we can deduce the channel gain given by
\begin{equation}\label{Gmu}
G_{\mu}=(1-\exp(-\eta\mu))+2 p_d,
\end{equation}
and the error rate given by
\begin{equation}\label{Qmu}
Q_{\mu}=\frac{Q_A(1-\exp(-\eta\mu))+p_d}{(1-\exp(-\eta\mu))+2 p_d}.
\end{equation}
From the definition of $\Delta$, we have
\begin{equation}\label{Delta}
\Delta=\frac{1}{2\eta}(\mu+O(\mu^2))
\end{equation}
The expression of $\mu$ is thus $\mu=2\eta\Delta$, showing that the average photon number in the signal pulse has to decrease proportionally to the channel transmission if one wants we keep $\Delta$ constant as a function of the distance. We see from Eq.~(\ref{Gmu}) that the rate is proportional to $\eta^2$ as long as $\eta$ is much greater than $p_d$.
Taking the same expression for $\eta$ and the same value for the parameters as in the previous section, we can plot the rate as a function of the distance and compare it with the result obtained in the single photon case, as displayed on figure (\ref{figureTauxDistance}). Here, for simplicity, we have supposed a perfect visibility $(V_A=1)$, and we obtain a limit distance of 90 km.

One counter-measure to PNS attacks is to introduce decoy states~\cite{Hwang2003,Wang2005,Lo2005prl}. Our goal is to show that decoy states can be combined with our protocol in order to reach rates that are close to those obtained with single photon pulses. We consider here the case of an asymptotic decoy state method as described in~\cite{Lo2005prl} which consists for Alice to send signal pulses with average photon number $\mu$ and to mix them with a fixed proportion of decoy states that can take all possible average photon number values between 0 and $\mu$. It has been proven in~\cite{Ma2005pra} that practical methods combining one weak decoy state and exploiting the vacuum (vacuum + weak decoy state protocol) can lead to results very close to the asymptotic protocol. Therefore, Alice and Bob can deduce precisely the portion of pulses having exactly $N$ photons. Thus any attempt to eavesdrop the channel unavoidably modifies the relative part of each $N$ photon states, which allows to detect Eve. The GLLP analysis~\cite{Gottesman2004,Lo2005prl} results in a simple expression for the secure rate given by
\begin{equation}\label{tauxdecoy}
R\geq q\{-G_\mu f(Q_\mu)h(Q_\mu)+G_1[1-h(Q_1)]\}
\end{equation}
$G_1$ and $Q_1$ are respectively the gain of the channel and the error rate for the one photon pulses, $f(Q_\mu)$ is the efficiency of the reconciliation algorithm. We consider the 2TS protocol with perfect visibility $(V_A=1)$ for simplicity. The first term in the parenthesis is the information that can be retrieved by the eavesdropper on the sifted key. The second term is the information available to Alice and Bob which can be obtained only from true single photon pulses. Considering the states at the output of Alice, given by Eq.~(\ref{psiAgen}), and taking into account the overall transmission $\eta$, we obtain the expression of the channel gain for the single photon pulses
\begin{equation}\label{G1}
G_1=\exp(-\mu)\mu\eta.
\end{equation}
The error rate in the case of single photon pulses is given by Eq.~(\ref{Qexp}). In order to be able to compare it with the single photon case, we take $q=1$, and we consider that Alice and Bob can perform perfect information extraction, which results in $f(Q_\mu)=1$. As previously, the overall transmission is given by $\eta=\eta_L\eta_D$. With the same expression and parameter values as previously and taking the optimum value $\mu=0.5$, we can plot the rate as a function of the distance and compare it with the result obtained in the single photon case.
The curves are displayed on figure (\ref{figureTauxDistance}).
Both of them decrease as $\eta$ with the distance as long as $\eta\gg p_d$. The cut-off distance is 250 km for the single photon protocol. It is slightly lower for the decoy state protocol (225 km) but comparable. This shows that our protocol combined with decoy states can reach secure distances that are comparable to those obtained with single photons. In addition, the use of low noise detectors such as SSPD allows to reach unprecedented long secure distances.

\section{Conclusion}
\label{sec:Conclusion}
We have given a complete security analysis of time-coding protocols where the bits are encoded in coherent single-photon pulses spanning successive time-slots~\cite{Debuisschert2004,Boucher2005}. To model the protocol, each time-slot is represented by a basis state in the Hilbert space. We first consider the original protocol~\cite{Debuisschert2004}. It is based on three time-slots and can be modeled with three orthogonal basis states. We have renamed it Three Time-Slots Protocol (3TS). The bits are encoded in single-photon states that are represented by non orthogonal superpositions of the basis states. Therefore, it is impossible for Eve to preserve the coherence without introducing errors.

The mathematical expression describing the protocol can be reformulated. This leads to another protocol requiring only two basis states and which has therefore been called Two Time-Slots protocol (2TS).
The bits are encoded on two orthogonal states corresponding to two successive time-slots. Additional pulses spanning the two adjacent time-slots and corresponding to a superposition of the two basis states are added in order to keep Eve from eavesdropping the key without introducing errors.
A complete analysis of this protocol is given. We calculate the Holevo quantity between Alice and Eve as a function of the quantum bit error rate (QBER) for different values of the interferometer visibility. We compare it to the Shannon information between Alice and Bob to deduce the advantage of information of Alice and Bob over Eve as a function of the QBER.
In the case of a perfect visibility of the interferometer, the results coincide with those obtained for the BB84 protocol, with a maximum allowed QBER of $11\%$.

We then analyse the security of the 3TS protocol. The security analysis is very similar to that of the 2TS protocol. This suggests an improved version of the 3TS protocol where additional pulses are sent in addition to the two initial pulses. Those pulses are described by a superposition of the two states encoding the key. Measuring the coherence between those two states allows Bob to ensure the security of the key. This protocol has been called Completed Three Time-Slots protocol (C3TS). The results are identical to those of the 2TS protocol. Then, we have shown that, although being less performant, the original 3TS protocol can still allow producing secure keys up to a QBER of the order of $5\%$.

We then compare the two protocols calculating the maximum secure distance in both cases. We consider a realistic implementation involving a standard single mode optical fibre and superconducting single photon detectors. We obtain secure distances with a cut-off of 253 km for the 2TS protocol or the C3TS protocol and 235 km for the original 3TS protocol in the case of a perfect visibility $(V_A=1)$.
Such secure distances have recently been demonstrated using superconducting single photon detectors and ultra low loss fibres~\cite{stucki2009high}.

In order to take into account the case of real experimental implementations involving faint pulses instead of single photon pulses, we consider the case of PNS attacks. The protocols are sensitive to those attacks, which imposes a quadratic decrease of the rate with the channel attenuation if one wants to preserve the security of the key. Introducing decoy states where Alice can modulate the average photon number in the pulses within a known proportion, Alice and Bob can rule out those attacks. As a result, the secure rate decreases linearly with the attenuation, which is similar to what is obtained with single photon pulses. The secure distance cut-off is 225 km, which is only slightly smaller to the cut-off distance obtained with single-photon pulses. As a result, our time-coding protocols combined with decoy states are able to give security distances exceeding 200 km. This feature, in addition to their easy implementation, makes them good candidates for field implementation of long distance QKD links.

\section*{Acknowledgements}
\noindent
We thank Philippe Grangier for helpful discussions and suggestions. We acknowledge support from the European Union under project SINPHONIA (contract number
NMP4-CT-2005-16433) and from the french Agence Nationale de la Recherche under project SEQURE (Grant No. ANR-07-SESU-011).

\section*{References}
\noindent

\bibliographystyle{unsrt}

\end{document}